\documentclass[
preprint,
showpacs,
preprintnumbers,
nofootinbib,
bibnotes,
amsmath,amssymb,
aps,
prd,
]{revtex4-1}

\usepackage[pdftex]{graphicx}
\usepackage{dcolumn}
\usepackage{bm}
\usepackage{color}
\usepackage{multirow}
\usepackage{mathrsfs}
\usepackage{amssymb}
\usepackage{amsmath}

\begin{document}

\preprint{INT-PUB-15-074}

\title{CPT, CP, and C transformations of fermions, and their consequences, in theories with B-L violation}

\author{Susan Gardner}
\email{\texttt{gardner@pa.uky.edu}}
\author{Xinshuai Yan}
\email{\texttt{xinshuai.yan@uky.edu}}
\affiliation{Department of Physics and Astronomy, University of Kentucky,
Lexington, Kentucky 40506-0055 USA}


\begin{abstract} 
We consider the transformation properties of fermions under the 
discrete symmetries CPT, CP, and C in the presence of B-L violation. 
We thus generalize the analysis of the known properties of Majorana neutrinos, probed via 
neutrinoless double beta decay, to include the case of Dirac fermions 
with B-L violation, which can be probed via neutron-antineutron oscillations. 
We show that the resulting CPT phase has implications for the interplay of 
neutron-antineutron oscillations with external 
fields and sources and consider
the differences in the Majorana dynamics of neutrinos and neutrons 
in the context of theories with self-conjugate isospin $I=0$ and $I=1/2$ fields.  
\end{abstract} 




\maketitle


\section{Introduction}

In theories with B-L violation the possibility of Majorana fermions, 
which are particles that are their own antiparticles, emerges. 
Such particles, as long known, 
have special transformation properties under the discrete symmetries 
CPT and CP, as well as  C~\cite{Kayser:1983wm,Kayser:1984ge,Carruthers:1971,Feinberg:1959}. 
Their observation 
would reveal the existence of dynamics beyond the
Standard Model (SM). 
B-L violation can appear in 
theories of quarks, that carry
baryon number B, and/or leptons, that carry lepton number L, 
though the possibility of Majorana neutrinos has had
the most scrutiny. 
This is because a crisp dichotomy can arise 
in the theoretical description of a 
massive neutrino: it 
can be either a Dirac or a Majorana particle, in that its mass can emerge from either
Dirac or Majorana mass terms. The neutrino mass 
could also emerge from mass terms of both types~\cite{Gelmini:1994az}, though even if the neutrino
were pseudo-Dirac~\cite{Wolfenstein:1981kw}, 
so that its Dirac mass would give a predominant contribution to its 
total mass, the mass eigenstates would be Majorana~\cite{Gribov:1968kq,Bilenky:1983wt}. 
Moreover, the observation of neutrinoless double beta decay~\cite{Elliott:2002xe} would
establish the existence of the Majorana neutrino because 
the existence of B-L violation would generate 
an effective Majorana mass term even if such a mass term
were not explicitly present~\cite{Schechter:1981bd}. 

The seminal papers of 
Kayser and Goldhaber~\cite{Kayser:1983wm} and Kayser~\cite{Kayser:1984ge} 
concern the analysis of the special CPT, CP, and C properties 
of Majorana fields and states and the implications of those properties for 
neutrinoless double beta decay. 
Earlier, Carruthers~\cite{Carruthers:1971}, as well as Feinberg and 
Weinberg~\cite{Feinberg:1959}, determined the existence 
of phase restrictions in the P and TC transformations, with 
 Carruthers~\cite{Carruthers:1971} analyzing the detailed properties of 
 particle self-conjugate multiplets. 
These works contain the implicit assumption that phase restrictions are 
associated with particle self-conjugate fields, or, alternatively, that 
B-L symmetry is only broken through the appearance of a Majorana field. 
In this paper we generalize this earlier work to the treatment of Dirac fields with 
B-L violation. In order to preserve the symmetry restrictions found in the Majorana case, 
we find that the phases associated with the action of the 
discrete symmetries on fermion fields must be restricted in order to address the 
symmetry transformations of fermion interactions with B-L violation. In the 
absence of B-L violation, the phases and thus the phase restrictions we describe 
have no physical impact, so that our considerations are specific to theories with B-L violation. 

Our analysis is pertinent to theories of both leptons and quarks 
with B-L violation, where we note that the possibility of B-L violation in the quark 
sector can be probed through
neutron-antineutron ($n$-${\bar n}$) oscillations.  
The $n$-${\bar n}$ system with B-L violation bears direct comparison to the possibility of a
pseudo-Dirac neutrino. 
We recall that, in the SM, the neutron and antineutron are Dirac fermions,
as are the quarks that comprise them, because 
quantum chromodynamics (QCD), the accepted theory
of the strong interactions, is a SU(3) gauge theory with a complex
fundamental representation~\cite{Brambilla:2014jmp}. The empirical success of the quark 
model, which explains the significant magnetic moments of the neutron and proton, 
suggests that the Dirac mass of the neutron dominates its measured mass. 
Indeed, the current empirical limit on the free $n$-${\bar n}$ oscillation time 
limits the Majorana mass to 
$\delta m = (\tau_{n \bar n})^{-1} \le 6 \times 10^{-29} \,{\hbox{MeV}}$ 
at 90\% C.L.~\cite{BaldoCeolin:1994jz}. We will find that 
the phase restrictions on the discrete symmetry transformations 
in the presence of B-L violation have important implications for the interplay
of $n$-${\bar n}$ oscillations with external fields and sources; in particular, 
they resolve the conflict between Refs.~\cite{Gardner:2014cma,Berezhiani:2015uya}. 
Generally, this interplay is key 
to improving the sensitivity of future experimental 
searches~\cite{Phillips:2014fgb,Milstead:2015toa}. 

Herewith we sketch an outline of the body of the paper. 
We begin, in Sec.~\ref{Major}, 
by recapping the Majorana phase constraints~\cite{Kayser:1983wm,Kayser:1984ge}
before building a Majorana field from Dirac fields in order to 
study the discrete symmetry transformations of the Dirac fields 
in the presence of B-L violation. We find, as a result, 
constraints on the phases in the discrete symmetry transformations of fermion fields. 
With these in place we then turn, in Sec.~\ref{DiracBmL}, 
to the CPT and CP transformation properties of 
B-L violating operators. Remarkably B-L violating operators can be constructed that are either
even or odd under CPT, even though all the operators are explicitly Lorentz invariant. 
The CPT phase restriction we derive changes the sign of the B-L violating operators under
CPT. With it in place, we find that the CPT-odd operators vanish upon use of fermion
anticommutation relations, so that the CPT theorem is respected~\cite{Greenberg:2002uu}. 
We consider the implications of these results in regards to the interplay of
$n$-${\bar n}$ oscillations with external fields and sources, as well as whether
their observation can connote a breaking of CP symmetry, in Sec.~\ref{Implic}. 
Noting the failure of locality in theories of self-conjugate fields with half-integral
isospin~\cite{Carruthers:prl,HuanLee:prl,fleming:prl,YSJin:plb}, 
we consider the compatibility of the appearance of B-L violation with the SM 
in Sec.~\ref{Origin}, 
before offering a final summary. 

\section{Majorana phase constraints}
\label{Major}

To determine the phase-factor restrictions on the discrete symmetry transformations 
that emerge in the Majorana case, we 
follow Refs.~\cite{Kayser:1983wm,Kayser:1984ge} and 
replace the Dirac field $\psi$ in the discrete symmetry transformations of 
Eqs.~(\ref{Cdef},\ref{Pdef},\ref{Tdef}) with a general Majorana field $\psi_m$, 
for which 
the plane-wave expansion 
is given by 
\begin{eqnarray}
\psi_{m}(x)= \int \frac{d^{3}\mathbf{p}}{(2\pi)^{3/2} \sqrt{2 E}}\sum_{s} \left\{ f(\mathbf{p}, s) u(\mathbf{p}, s) e^{-ip\cdot x}
+ \lambda f^\dagger(\mathbf{p}, s) v(\mathbf{p}, s) e^{ip\cdot x}
\right\}\,. 
\label{plmajor}
\end{eqnarray}
We note that $f^{\dagger}$ and $f$ denote the creation and annihilation operators 
for the Majorana particle of interest. 
The unimodular 
quantity $\lambda$ is called a creation phase factor; it may be present, in general, 
and can be chosen arbitrarily. We refer the reader to Appendix \ref{CPTdef} for 
a summary of our definitions, conventions, and other useful basic results. 

Noting the C transformation
\begin{eqnarray}
\mathbf{C}\psi_{m}(x)\mathbf{C}^{-1}&=&i\eta_{c}\gamma^{2}\psi^{\ast}_{m}(x) 
\end{eqnarray}
and applying the Majorana relation, 
\begin{equation}
i\gamma^{2}\psi^{\ast}_{m}(x)=\lambda^{\ast}\psi_{m}(x),
\label{majorcon}
\end{equation}
yields 
\begin{eqnarray}
\mathbf{C}\psi_{m}(x)\mathbf{C}^{-1}&=&\eta_{c}\lambda^{\ast}\psi_{m}(x) 
\end{eqnarray}
and thus 
\begin{eqnarray}
\mathbf{C}f(\mathbf{p},s)\mathbf{C}^{-1}&=&\eta_{c}\lambda^{\ast}f(\mathbf{p},s) \,,\\
\mathbf{C}f^{\dagger}(\mathbf{p},s)\mathbf{C}^{-1}&=&\eta_{c}\lambda^{\ast}f^{\dagger}(\mathbf{p},s) \,.
\end{eqnarray}
Since $\mathbf{C}$ is a unitary operator, taking the Hermitian conjugate of either relation 
reveals that $\eta^{\ast}_{c}\lambda$ is real. 
Noting the CP transformation 
\begin{eqnarray}
\mathbf{CP}\psi_{m}(t,\mathbf{x})(\mathbf{CP})^{-1}&=&i\eta_{p}\eta_{c}\gamma^{0}\gamma^{2}\psi^{\ast}_{m}(t,-\mathbf{x}) 
\end{eqnarray}
and Eq.~(\ref{majorcon}) yields 
\begin{eqnarray}
\mathbf{CP}\psi_{m}(t,\mathbf{x})(\mathbf{CP})^{-1}&=&\eta_{p}\eta_{c}\lambda^{\ast}\gamma^{0}\psi_{m}(t,-\mathbf{x})\,
\end{eqnarray}
and thus
\begin{eqnarray}
\mathbf{CP}f(\mathbf{p},s)(\mathbf{CP})^{-1}&=&\eta_{c}\eta_{p}\lambda^{\ast}f(-\mathbf{p},s)\,,\\
\mathbf{CP}f^{\dagger}(\mathbf{p},s)(\mathbf{CP})^{-1}&=&-\eta_{c}\eta_{p}\lambda^{\ast}f^{\dagger}(-\mathbf{p},s)\,. 
\end{eqnarray}
Since $\mathbf{CP}$ is a unitary operator, taking the Hermitian conjugate of either relation
shows that $\eta^{\ast}_{p}\eta^{\ast}_{c}\lambda$ must be imaginary. 
We have already established that $\eta^{\ast}_{c}\lambda$ is real, so that 
$\eta^{\ast}_{p}$ itself must be imaginary. Under T we have 
\begin{eqnarray}
\mathbf{T}\psi_{m}(t,\mathbf{x})\mathbf{T}^{-1}&=&\eta_{t}\gamma^{1}\gamma^{3}\psi_{m}(-t,\mathbf{x}), 
\end{eqnarray}
which yields 
\begin{eqnarray}
\mathbf{T}f(\mathbf{p},s)\mathbf{T}^{-1}&=& s \eta_{t}  f(-\mathbf{p},-s), \\
\mathbf{T}f^{\dagger}(\mathbf{p},s)\mathbf{T}^{-1}&=& s \eta_{t}\lambda^{2}  f^{\dagger}(-\mathbf{p},-s).
\end{eqnarray}
Since $\textbf{T}$ is an antiunitary operator, we write 
$\textbf{T} = K U_{t}$, where $U_{t}$ is a unitarity operator and $K$ denotes complex
conjugation. Then taking the Hermitian conjugate of either relation
shows that $\eta_{t}\lambda$ must be real.  Finally 
we note the CPT transformation of $\psi_{m}$ 
\begin{eqnarray}
\mathbf{CPT}\psi_{m}(x)(\mathbf{CPT})^{-1}
=-\eta_{c}\eta_{p}\eta_{t}\gamma^{5}\psi^{\ast}_{m}(-x)\,,
\end{eqnarray}
with $\gamma^{5}\equiv i\gamma^{0}\gamma^{1}\gamma^{2}\gamma^{3}$, which yields,
\begin{eqnarray}
\xi f(\mathbf{p}, s)\xi^{-1}&=&s\lambda^{\ast}\eta_{c}\eta_{p}\eta_{t}f(\mathbf{p},-s)\,, 
\\
\xi f^{\dagger}(\mathbf{p}, s)\xi^{-1}&=&-s\lambda\eta_{c}\eta_{p}\eta_{t}
f^{\dagger}(\mathbf{p},-s)\,,
\end{eqnarray}
where we employ $\textbf{CPT}\equiv \xi$. 
Since $\xi$ is an antiunitary operator, we write 
$\xi = K U_{cpt}$, where $U_{cpt}$ denotes a unitarity operator.
Consequently, taking the Hermitian conjugate of either relation reveals 
that $\eta_c \eta_p \eta_t$ is pure imaginary. Since we have already established
that $\eta_p$ is imaginary, we see that $\eta_c \eta_t$ must also be real --- and note that
just this emerges from the analysis of the TC transformation as well. 
In contrast, the combination $\eta_c \eta_p$ itself is unconstrained. 
In summary, we have found all the restrictions on the phases that appear in 
C, P, T, and combinations thereof, 
and our results are equivalent to 
those in Refs.~\cite{Kayser:1983wm,Kayser:1984ge,Carruthers:1971}. 

We now turn to the particular case of a 
Majorana field that is constructed from Dirac fields. 
Given Eqs.~(\ref{Cdef},\ref{Pdef},\ref{Tdef}), 
the existence of phase restrictions in the application of C, P, and T to Dirac fields
themselves may already be self-evident. However, we confirm this through explicit calculation. 
Thus we build 
$\psi_{m}$ from the linear combination 
$a\psi +b \textbf{C}\psi\textbf{C}^{-1}$ in which 
$a$ and $b$ are complex numbers to be determined. 
Under C, $\psi_{m}$ becomes
\begin{eqnarray}
\textbf{C}\psi_{m}\textbf{C}^{-1} &=& 
\frac{b}{a}(a\psi+\frac{a^2}{b}\textbf{C}\psi\textbf{C}^{-1})\,. \nonumber 
\end{eqnarray}
Since $\psi_{m}$ is a Majorana field, $\textbf{C}\psi_{m}\textbf{C}^{-1} \propto \psi_{m}$,
yielding the condition $a^2 = b^2 $, i.e., $a=\pm b$. 
After imposing a normalization condition on $\psi_{m}$, we find 
\begin{eqnarray}
\psi_{m\pm}(x)=\frac{1}{\sqrt{2}}(\psi(x) \pm \textbf{C}\psi(x)\textbf{C}^{-1})\,,
\end{eqnarray}
which has the plane-wave expansion 
\begin{eqnarray}
&&\psi_{m\pm}=\int \frac{d^{3}\mathbf{p}}{(2\pi)^{3/2} \sqrt{2 E}}\sum_{s} \{ \frac{1}{\sqrt{2}}[b(\textbf{p},s)\pm\eta_{c}d(\textbf{p},s)]u(\mathbf{p}, s) e^{-ip\cdot x} \nonumber \\
&&\ \ \ \ \ \ \ \ \ \ \ \ \ \ \ \ \ \ \ \ \ \ \ \ \ \ \ \ \ \ \ \ \ + \frac{1}{\sqrt{2}}[d^{\dagger}(\textbf{p},s)\pm\eta_{c}b^{\dagger}(\textbf{p},s)] v(\mathbf{p}, s) e^{ip\cdot x}\}.
\end{eqnarray} 
Comparing with Eq.~(\ref{plmajor}), we define 
\begin{eqnarray}
w_{m\pm}(\textbf{p},s)&\equiv &\frac{1}{\sqrt{2}}[b(\textbf{p},s)\pm\eta_{c}d(\textbf{p},s)],
\end{eqnarray}
and observe that the second term can be written as
\begin{eqnarray}
\frac{1}{\sqrt{2}}(d^{\dagger}_{s}(\textbf{p})\pm\eta_{c}b^{\dagger}_{s}(\textbf{p}))=\pm\eta_{c}w^{\dagger}_{m\pm}(s,\textbf{p}), 
\end{eqnarray}
so that we can rewrite $\psi_{m\pm}$ in a simple way
\begin{eqnarray}
\psi_{m\pm}(x)= \int \frac{d^{3}\mathbf{p}}{(2\pi)^{3/2} \sqrt{2 E}}\sum_{s} \left\{ w_{\pm}(\mathbf{p}, s) u(\mathbf{p}, s) e^{-ip\cdot x}
\pm\eta_{c} w_{\pm}^\dagger(\mathbf{p}, s) v(\mathbf{p}, s) e^{ip\cdot x}
\right\}.
\end{eqnarray}
Comparing with Eq.~(\ref{plmajor}), we find that 
$\lambda$ is no longer arbitrary; rather, $\lambda=\pm\eta_{c}$.
Since $\psi_{m\pm}$ is a Majorana field, our earlier lines of reasoning, as well 
as our conclusions, should still apply. 
Note, e.g., that applying the C transformation to $\psi_{m\pm}$ yields
\begin{eqnarray}
\mathbf{C}\psi_{m\pm}(x)\mathbf{C}^{-1}=\frac{1}{\sqrt{2}}[(\eta_{c}i\gamma^{2})\psi^{\ast}(x)\pm \psi(x)]=\pm\psi_{m\pm} \,, 
\end{eqnarray}
which is automatically consistent with our earlier conclusion that 
$\eta_c^\ast \lambda$ is real, since $\lambda = \pm \eta_c$. 
Turning to the explicit CP and CPT transformation properties of $\psi_{m\pm}$ we confirm our
earlier results that both $\eta_c^*\eta_p^*\lambda$ (or $\eta_p$) 
and $\eta_c \eta_p \eta_t$ are imaginary
--- and thus that 
$\eta_c \eta_t$ is real. 
Interestingly, the study of T and CT (or TC) transformations lead to no further phase 
restrictions. 
Under T, $\psi_{m\pm}$ becomes
\begin{eqnarray}
\textbf{T}\psi_{m\pm}(t,\mathbf{x})\textbf{T}^{-1}&=&\frac{1}{\sqrt{2}}\{\eta_{t}\gamma^{1}\gamma^{3}\psi(-t,\mathbf{x}) \pm (\eta_{c}\eta_{t})^{\ast}(i\gamma^{2})^{\ast}\gamma^{1}\gamma^{3}\psi^{\ast}(-t,\mathbf{x})\}\,,
\end{eqnarray}
but noting Eq.~(\ref{Tdef}) this should be equivalent to
\begin{eqnarray}
\eta_{t}\gamma^{1}\gamma^{3}\psi_{m\pm}(-t,\mathbf{x})=\frac{1}{\sqrt{2}}\{\eta_{t}\gamma^{1}\gamma^{3}\psi(-t,\mathbf{x})\pm i\eta_{t}\eta_{c}\gamma^{1}\gamma^{3}\gamma^{2}\psi^{\ast}(-t,\mathbf{x})\}\,; 
\end{eqnarray} 
and we conclude that $\eta_{c}\eta_{t}$ is real. 
Upon applying CT (or TC) to $\psi_{m\pm}$ we find just the same constraint: that 
$\eta_{c}\eta_{t}$ must be real.

In summary, we have found that in order 
to preserve the phase restrictions
found in the Majorana case, the phases in the discrete symmetry transformations of 
fermion fields must themselves be restricted. Specifically we have found that 
$\eta_p$ must be imaginary 
and that the combination $\eta_c \eta_t$ must be real. As a result, we find that 
$\mathbf{P}^2 \psi(x) \mathbf{P}^{-2}  = - \psi(x)$. Furthermore, we find that
although $\eta_c\eta_p\eta_t$ is pure imaginary the combination 
$\eta_c\eta_p$ is unconstrained.  

Before proceeding we note that the phase restrictions 
we have found are not restricted to our particular choice of gamma matrix representation
and that certain aspects thereof apply to the transformations of 
two-component (Majorana) fields as well. 
For definiteness we consider representations in which 
$(\gamma^\mu)^\dagger = \gamma^0 \gamma^\mu \gamma^0$ is satisfied, so that
Eq.~(\ref{Cdef}) holds~\cite{Branco:1999fs}. This subset of possible representations 
includes the Weyl and Majorana representations as well, so that 
our choice spans 
all the commonly used ones.
Moreover, unitary transformations exist that connect all the representations for which 
Eq.~(\ref{Cdef}) holds~\cite{MPbook}. For completeness, we present the particular
phase restrictions associated with the discrete-symmetry transformations 
of two-component Majorana fields in Appendix \ref{CPT2spinor}. 

\section{Theories of Dirac fermions with B-L violation}
\label{DiracBmL}

We now turn to the implications of the phase constraints we have discussed and
begin by considering the discrete symmetry transformations of various 
B-L violating operators with Dirac fields, for which the prototypical example
is 
\begin{equation}
\psi^T C \psi + \hbox{h.c.}\,,
\label{nnbarop}
\end{equation}
where ``h.c.'' denotes the Hermitian conjugate. 
Note that $C$ satisfies $(\sigma^{\mu\nu})^{\rm T} C = -C \sigma^{\mu\nu}$, so that the 
construction of Eq.~(\ref{nnbarop}) is automatically Lorentz invariant. 
In what follows 
we work at energies far below the scale of B-L breaking; indeed, we work at 
sufficiently low-energy scales that we suppose 
the Dirac field $\psi$ can be regarded as  elementary. 
Moreover, since the primary use of such operators
will be in theories of neutron-antineutron oscillations, or in theories
of pseudo-Dirac neutrinos, we assume that the mass associated with the fermion field 
is dominated by its Dirac mass; this reduces the list of possible non-trivial operators 
that can appear. In what follows we enumerate all the lowest
mass dimension B-L violating operators with Lorentz structures that span the possible bilinear 
covariants and discuss their transformation properties under CPT, as well as CP. 
We do not include operators with derivatives on the fermion field
operators because the free-particle Dirac equation can be used to bring them
to the form of those we do include. Thus we consider operators ${\cal O}_i$, namely, 
\begin{eqnarray}
&{\cal O}_1 =
\psi^T C \psi + {\rm h.c.} \, \quad &\stackrel{\mathbf{CPT}}{\Longrightarrow} 
-(\eta_c \eta_p \eta_t)^2     \,,
\\
&{\cal O}_2 =
\psi^T C \gamma_5 \psi + {\rm h.c.} \, \quad &\stackrel{\mathbf{CPT}}{\Longrightarrow}  -(\eta_c \eta_p \eta_t)^2    \,, \\
&{\cal O}_3 = 
\psi^T C \gamma^\mu \psi \, \partial^\nu F_{\mu\nu} + {\rm h.c.} \, \quad &\stackrel{\mathbf{CPT}}{\Longrightarrow}  +(\eta_c \eta_p \eta_t)^2     \,, 
\label{wrong1} \\
&{\cal O}_4 = 
\psi^T C \gamma^\mu \gamma_5 \psi \, \partial^\nu F_{\mu\nu} 
 + {\rm h.c.} \,  \quad &\stackrel{\mathbf{CPT}}{\Longrightarrow}  -(\eta_c \eta_p \eta_t)^2    \,, \\
&{\cal O}_5 =
\psi^T C \sigma_{\mu \nu} \psi\, F^{\mu \nu} + {\rm h.c.} \, \quad &\stackrel{\mathbf{CPT}}{\Longrightarrow}  +(\eta_c \eta_p \eta_t)^2     \,, 
\label{wrong2} \\
&{\cal O}_6 =
\psi^T C \sigma_{\mu \nu} \gamma_5 \psi\, F^{\mu \nu} + {\rm h.c.} \, \quad &\stackrel{\mathbf{CPT}}{\Longrightarrow}  +(\eta_c \eta_p \eta_t)^2     \,, 
\label{wrong3} 
\end{eqnarray}
where we have included the axial tensor operator ${\cal O}_6$ even if not strictly necessary, 
and we have reported the phase factor for the 
transformation of each operator under CPT as well. 
Note that we have included 
the electromagnetic field strength tensor $F^{\mu \nu}$ and its source
as needed to make the B-L violating operators transform as Lorentz scalars. 
Remarkably, the set of operators ${\cal O}_i$ do not transform under 
CPT with a definite sign, and the phase constraints we have derived in 
Sec.~\ref{Major}, that $(\eta_c \eta_p \eta_t)^2=-1$, only serves to flip the
sign of each eigenvalue. The existence of CPT-odd operators that are Lorentz scalar 
is in apparent contradiction with the CPT theorem~\cite{Greenberg:2002uu}, 
which asserts that CPT breaking implies that Lorentz symmetry is broken also. 
Nevertheless, the theorem remains secure, because, as we shall show, the operators
of Eqs.~(\ref{wrong1},\ref{wrong2},\ref{wrong3}) vanish once the anticommuting
nature of fermion fields is taken into account. This anticommuting behavior is
implicit to the determination of the transformation of the Dirac bilinears under
C and CPT and is not an additional assumption.
That only the operators of 
Eqs.~(\ref{wrong1},\ref{wrong2},\ref{wrong3}) vanish outright 
speaks to the key nature of the phase
constraint $(\eta_c \eta_p \eta_t)^2=-1$ in making 
theories with B-L violation consistent with the tenets of quantum field theory. 

The idea that the operators in Eqs.~(\ref{wrong1},\ref{wrong2},\ref{wrong3}) 
should have no effect has been 
discussed in particular contexts, though never from the viewpoint of their wrong CPT. 
For example, the vector, tensor, and axial tensor electromagnetic form factors of Majorana
neutrinos have been shown to vanish~\cite{Nieves:1981zt,Schechter:1981hw,Kayser:1982br,Shrock:1982sc,Li:1981um,Davidson:2005cs}, and we refer the reader to the succinct treatment of 
Ref.~\cite{MPbook}.  
Similarly, in the phenomenology of flavor-spin neutrino oscillations, 
the flavor-diagonal $\nu$ transition magnetic moment has been noted to 
vanish~\cite{Okun:1986na,Okun:1986hi,Lim:1987tk}. 
We now establish that the operators of 
Eqs.~(\ref{wrong1},\ref{wrong2},\ref{wrong3}) vanish regardless of whether
Majorana or Dirac fields are employed.

\subsection{CPT-odd operators with Majorana fields}

In the case of Majorana fields, for which Eq.~(\ref{majorcon}) holds, 
we can immediately show that the operators of 
Eqs.~(\ref{wrong1},\ref{wrong2},\ref{wrong3}) --- and only these of our list --- 
vanish identically, and that
this follows from the anticommuting nature of fermion fields. 
We note that Eq.~(\ref{majorcon}) can be rewritten as any of 
$\psi_m^T C = \lambda {\bar \psi}_m$, $C^\dagger \psi_m^\ast 
= \lambda^\ast \gamma^0 \psi_m$, 
$\psi_m^\dagger C^\dagger = -\lambda^\ast \psi_m^T \gamma^0$, and 
$C \psi_m = - \lambda \gamma^0 \psi_m^\ast$. Thus ${\cal O}_1$, e.g., can be rewritten 
as $(\lambda + \lambda^\ast) {\bar \psi}_m \psi_m$ or 
$-(\lambda + \lambda^\ast) \psi_m^T {\bar \psi}_m^T$, but these are equal because 
${\bar \psi}_m \psi_m= - \psi_m^T {\bar \psi}_m^T$. Therefore ${\cal O}_1$ need not
vanish. Similarly for ${\cal O}_2$ we have 
$(\lambda - \lambda^\ast) {\bar \psi}_m \gamma_5 \psi_m$, or 
$-(\lambda - \lambda^\ast) \psi_m^T \gamma_5 {\bar \psi}_m^T$, 
and thus ${\cal O}_2$ also
need not vanish. Noting that $C \gamma^\mu = -\gamma^{\mu \,T} C$ we see, however, 
that ${\cal O}_3 = (\lambda + \lambda^\ast) {\bar \psi}_m \gamma^\mu \psi_m j_\mu
= (\lambda + \lambda^\ast) {\psi}_m^T \gamma^{\mu\,T}  {\bar \psi}_m^T j_\mu$, 
with $j_\mu \equiv \partial^\nu F_{\mu\nu}$,  and thus ${\cal O}_3$ vanishes. 
In contrast, we have 
that ${\cal O}_4 = (\lambda - \lambda^\ast) {\bar \psi}_m \gamma^\mu \gamma_5 \psi_m 
j_\mu
= - (\lambda - \lambda^\ast) {\psi}_m^T \gamma_5 \gamma^{\mu\,T} 
{\bar \psi}_m^T j_\mu$, 
and we conclude that ${\cal O}_4$ can be nonzero. Finally, since 
$(\sigma^{\mu \nu})^T C \gamma^\mu = - C \sigma^{\mu \nu}$, 
we have that 
${\cal O}_5 = (\lambda + \lambda^\ast) {\bar \psi}_m \sigma^{\mu \nu} \psi_m F_{\mu \nu}
= (\lambda + \lambda^\ast) {\psi}_m^T (\sigma^{\mu \nu})^T {\bar \psi}_m^T F_{\mu \nu}
$, as well as 
${\cal O}_6 = (\lambda - \lambda^\ast) {\bar \psi}_m \sigma^{\mu \nu} \gamma_5\psi_m 
F_{\mu \nu}
= (\lambda - \lambda^\ast) {\psi}_m^T \gamma_5 (\sigma^{\mu \nu})^T 
{\bar \psi}_m^T F_{\mu \nu}$. 
We see that both ${\cal O}_5$ and ${\cal O}_6$ vanish as well. Thus 
we have proven what we set out to show.

\subsection{CPT-odd operators with Dirac fields}

In the case of Dirac fields, for which Eq.~(\ref{majorcon}) does not hold, 
a similarly ready proof that the operators of Eqs.~(\ref{wrong1},\ref{wrong2},\ref{wrong3}) 
vanish 
is not available. In this case we evaluate
the operators explicitly by postulating that the field operators 
satisfy equal-time anticommutation relations and expanding them
in 
the free-particle, plane-wave expansion of Eq.~(\ref{free}). We then 
immediately find that ${\cal O}_5$ and ${\cal O}_6$~\cite{Gardner:2014cma}, as well as 
${\cal O}_3$, vanish due to the anticommuting nature of fermion fields. 
Since our demonstration assumes that the fermion is both free and point-like, 
we now turn to ways in which we can make it more general, 
considering the conditions under which 
we can extend it 
to the case of bound particles, 
as well as to that of strongly bound composite particles. 
We would like our conclusions to be pertinent to $n-{\bar n}$ oscillations, for both
free and bound neutrons. 

In the case that the particle is loosely bound, e.g., 
the effect of the ``wrong CPT'' operators is still zero because 
the loosely bound state can be regarded 
as a linear superposition of free states of momentum $\textbf{k}$, weighted by 
its wave function~\cite{Peskin:1995ev}. 
Since the wrong CPT operators vanish for free states, then the operators involving
such loosely bound particles will also. We note that since the binding energies of
neutrons in large nuclei are no more than $\sim8$ MeV per particle, our argument should be 
sufficient to conclude that Eqs.~(\ref{wrong1},\ref{wrong2},\ref{wrong3}) do not operate
for bound neutrons.

An interesting question may be what happens 
if the fermion is actually a strongly bound composite particle, such as the
neutron itself. We have explored this in the particular case of $n-{\bar n}$ 
oscillations using the M.I.T. bag model~\cite{Chodos:1974je,Chodos:1974pn}, 
following the analysis of Ref.~\cite{Rao:1982gt}. 
Since the quarks within the bag are free, an expansion of the quark fields 
in single-particle modes analogous to 
Eq.~(\ref{free}) exists~\cite{Chodos:1974pn}, 
suggesting that the results of our earlier analysis at
the nucleon level should be pertinent here as well. Indeed an explicit calculation 
of the transition  matrix element $\langle {\bar n} | O_1 | n \rangle$ 
using the $O_1$ operator of Ref.~\cite{Rao:1982gt}
with the substitution of $u^{T\,\alpha}_{\chi 1}C\sigma^{\mu\nu}u^{\beta}_{\chi 1}F_{\mu\nu}$ 
for $u^{T\,\alpha}_{\chi 1}C u^{\beta}_{\chi 1}$ yields zero. 
In what follows we assume that the operators of 
Eqs.~(\ref{wrong1},\ref{wrong2},\ref{wrong3}) do indeed vanish if 
Lorentz symmetry is not broken. As an aside, we 
note that an explicit proof of the CPT theorem
within confining theories is still lacking~\cite{Kostelecky:1997hw}.

\subsection{CP transformation properties}

We now turn to the analysis of the CP properties of the surviving B-L violating operators, 
finding 
\begin{eqnarray}
&{\cal O}_1 =
\psi^T C \psi + {\rm h.c.} \, \quad &\stackrel{\mathbf{CP}}{\Longrightarrow} 
-(\eta_c \eta_p )^2     \,,
\label{cpO1}
\\
&{\cal O}_2 =
\psi^T C \gamma_5 \psi + {\rm h.c.} \, \quad &\stackrel{\mathbf{CP}}{\Longrightarrow}  -(\eta_c \eta_p )^2    \,, 
\label{cpO2} 
\\
&{\cal O}_4 = 
\psi^T C \gamma^\mu \gamma_5 \psi \, \partial^\nu F_{\mu\nu} 
 + {\rm h.c.} \,  \quad &\stackrel{\mathbf{CP}}{\Longrightarrow}  -(\eta_c \eta_p )^2    \,, 
\end{eqnarray}
where we have left the phase dependence explicit. 
Noting 
 our earlier determined phase constraint that $\eta_p^2=-1$, we see, nevertheless, that the 
CP transformation properties of the operators are not definite --- rather, 
they are given by $\eta_c^2$, where $\eta_c$ is not determined. 
Explicit examples of the indeterminate nature of the CP transformation, illustrated
through the phase rotation $\psi \to \psi'=e^{i\theta}\psi$, 
can be found in Ref.~\cite{Fujikawa:2015iia}. The noted phase rotation has 
the effect of changing $\eta_c \to e^{2i\theta}\eta_c$, 
$\eta_t \to e^{-2i\theta}\eta_t$, 
with $\eta_p$ unchanged, under $\psi \to \psi'$ 
in the C, T, and P transformations, respectively. We emphasize that the indeterminacy
arises from that in $\eta_c^2$ and thus emerges generally for B-L violating operators. 
In Ref.~\cite{Fujikawa:2015iia} $\eta_c=\eta_p=1$ and $\eta_t=i$ throughout, and 
although these choices are consistent with the phase constraint we have found 
for the CPT transformation, they are not consistent with 
the phase constraints we have found for P and TC, though this does not 
impact their conclusion regarding the indeterminacy of CP. 
If $\eta_c^2$ were set to $-1$, then Eq.~(\ref{cpO1}) gives the result reported
in Ref.~\cite{Berezhiani:2015uya}. 
We argue on physical grounds that the observation of $n-{\bar n}$ oscillations
cannot itself constitute evidence of CP violation in the following section. 

\section{Implications of the CPT and CP phases}
\label{Implic}

In this section we consider the consequences of the CPT and CP transformation
properties we have determined in previous sections, particularly in regards to 
their implications for the interplay of the appearance of $n-{\bar n}$ oscillations with 
external magnetic fields. It has long been thought that 
experimental searches for free $n-{\bar n}$ oscillations must be performed 
in a high-vacuum, low-magnetic-field environment, because 
the energy of a neutron and antineutron generally ceases to be the same 
in the presence of
matter or magnetic fields, suppressing $n-{\bar n}$ 
oscillations~\cite{Mohapatra:1980de,Cowsik:1980np}. 
However, if a $n-{\bar n}$ transition could connect a neutron and antineutron of opposite spin, 
then CPT invariance would guarantee that those states would be of the 
same energy in a magnetic field --- and eliminating the magnetic field
would no longer be necessary. 
In Ref.~\cite{Gardner:2014cma} 
it was argued that spin-dependent SM effects involving
transverse magnetic fields could, in effect, realize $n-{\bar n}$ transitions in 
which the particle spin flips and thus accomplish this goal. However, this
conclusion is sensitive to 
the CPT phase constraint we have discussed. 
To illustrate, we revisit the example analyzed in Ref.~\cite{Gardner:2014cma}: 
a neutron at rest 
that can oscillate to 
an antineutron is in a static magnetic field $\mathbf{B}_0$ 
and to which a static transverse field ${\bf B}_1$ is suddenly applied at $t=0$. 
Noting that $\mathbf{B}_0$ fixes the spin quantization axis and 
defining $\omega_0 \equiv - \mu_n B_0$ and $\omega_1 \equiv - \mu_n B_1$, the  
Hamiltonian matrix in the $|n(+)\rangle$, $|{\bar n}(+)\rangle$, 
$|n(-)\rangle$, $|{\bar n}(-)\rangle$ basis at $t>0$ is of form 
\begin{equation}
{\cal H} = 
\left(\begin{array}{cccc}
M + \omega_0 & \delta & \omega_1 & 0 \\
\delta & M - \omega_0 & 0 & -\omega_1 \\
\omega_1  & 0 & M - \omega_0 & -\delta \eta_{cpt}^2 \\
0 & -\omega_1 & -\delta \eta_{cpt}^2 & M + \omega_0  
\end{array}\right) \,,
\end{equation}
where $M$ is the neutron mass and 
$\delta$, which 
is real in this example, denotes a $n(+)\to{\bar n}(+)$ transition matrix element. The 
other signs are fixed by Hermiticity and CPT invariance. We
have now explicitly included the dependence of the B-L violating operator on the 
phase of the CPT transformation, namely, $\eta_{cpt} \equiv \eta_c\eta_p\eta_t$. 
In Ref.~\cite{Gardner:2014cma} the phase $\eta_{cpt}$ was 
set to unity; in this work we have, rather, established that $\eta_{cpt}^2 =-1$. 

In Ref.~\cite{Gardner:2014cma} the unpolarized $n$-${\bar n}$ transition probability 
was found to be, noting $|\delta| \ll |\omega_0|\,, |\omega_1|$, 
\begin{eqnarray}
{\cal P}_{n\, \to {\bar n}} (t)  &=& \delta^2\Bigg[
\frac{\omega_1^2 t^2}{\omega_0^2 + \omega_1^2} + 
\frac{\omega_0^2}{(\omega_0^2 + \omega_1^2)^2} \sin^2 (t\sqrt{\omega_0^2 + \omega_1^2})
 \nonumber \\
&+&  \frac{\omega_0^2 \omega_1^2 t}{(\omega_0^2 + \omega_1^2)^{5/2}}
\left(1 - \sin \left(2t\sqrt{\omega_0^2 + \omega_1^2}\right)\right) \Bigg] 
+ {\cal O}(\delta^3) \,,
\end{eqnarray}
where if $|\omega_0| \sim |\omega_1|$ 
the first term is of ${\cal O}(1)$ in magnetic fields 
--- and thus the 
quenching previously noted no longer appears. 
However, the exact eigenvalues at $t>0$ are 
\begin{eqnarray}
E_1 &=& M - \sqrt{\omega_0^2 + (\delta - \omega_1)^2} 
\,, \quad \nonumber \nonumber \\
E_2 &=& M + \sqrt{\omega_0^2 + (\delta - \omega_1)^2} 
\,, \quad \nonumber \nonumber \\
E_3 &=& M - \sqrt{\omega_0^2 + (\delta + \omega_1)^2} 
\,, \quad \nonumber \nonumber \\
E_4 &=& M + \sqrt{\omega_0^2 + (\delta + \omega_1)^2} \,. 
\label{wrongE}
\end{eqnarray}
As pointed out in Refs.~\cite{volo,Berezhiani:2015uya}, 
this is incompatible with rotational invariance because the
eigenenergies do not depend on the magnitude of the total 
magnetic field $|\mathbf{B}|$ alone. 
However, once we have included the needed phase $\eta_{cpt}^2=-1$, we then find that the 
energy eigenvalues at $t>0$ do indeed depend on $|\mathbf{B}|$, as needed 
by rotational invariance~\cite{volo,Berezhiani:2015uya}, recovering the form 
found in Ref.~\cite{Berezhiani:2015uya}, and 
that $n(+) \to {\bar n}(-)$ and $n(-) \to {\bar n}(+)$ transitions no longer
occur.  As a result, $n{\bar n}$ transitions are 
quenched irrespective of the presence of transverse magnetic fields. We note that 
employing time-dependent magnetic fields in the manner familiar from 
the theory of magnetic resonance~\cite{Rabi,CTDL}, as discussed in Ref.~\cite{Gardner:2014cma}, 
does not change this conclusion --- the time-dependent case, upon
a change of variable, resembles 
the static case we have already analyzed. Finally, then, the failure
of rotational invariance in Eq.~(\ref{wrongE})~\cite{Gardner:2014cma} 
is a consequence of 
the inadvertent use of a Hamiltonian matrix in 
which the $n-{\bar n}$ transition operator broke
CPT and hence Lorentz invariance; this is redressed through the inclusion of
the phase $\eta_{cpt}$. 

We now turn to the possibility of CP violation in free
$n-{\bar n}$ oscillations in the absence of external fields, for which 
the $n-{\bar n}$ transition probability is controlled by 
$|\delta|^2$~\cite{Mohapatra:1980de}.  Referring
to Eqs.~(\ref{cpO1},\ref{cpO2}), though only Eq.~(\ref{cpO1}) operates~\cite{Gardner:2014cma}, 
we see that the probability transforms as $|\eta_c|^2=1$. Thus even if $\delta$ does
not have definite CP its associated observable is CP even. 
Consequently the observation of free $n-{\bar n}$ oscillations 
cannot itself constitute a CP-violating effect. 
This is in contradistinction to the case of a permanent electric-dipole moment (EDM) $d$, for
which the low-energy Hamiltonian for a particle with spin $\mathbf{S}$ is 
\begin{equation}
{\cal H} = - \frac{\mu}{S} {\mathbf S}\cdot {\mathbf B} - \frac{d}{S} {\mathbf S}\cdot {\mathbf E}
\,.
\end{equation}
Here a nonzero value of $d$ generates an observable CP-violating effect, even if it is 
generated by a single operator, because 
the spin-state energy splitting generated by the $\mu$-term in a nonzero magnetic
field changes upon the reversal 
of an applied electric field. 

We conclude this section by noting that 
despite the failure of the specific method proposed in Ref.~\cite{Gardner:2014cma}, 
spin-dependent effects could well prove key to realizing 
$n-{\bar n}$ oscillations. In particular, the $n-{\bar n}$ transition operator 
\begin{equation}
{\cal O}_4 = \psi^T C \gamma^\mu \gamma_5 \psi \, \partial^\nu F_{\mu\nu} + {\rm h.c.}
\end{equation}
couples states of the same energy in a magnetic field, so that, in effect, 
$n(+)\to {\bar n}(-)$ can occur directly because the interaction with an external source, such as 
an electron beam, 
flips the spin. This is concomitant with 
the study of the crossed process 
$n(p_1,s_1) + n(p_2,s_2) \to \gamma^* (k)$, 
for which only $L=1$
and $S=1$ is allowed in the initial state via angular momentum conservation and
Fermi statistics~\cite{Berezhiani:2015uya}. 
As a result, this particular operator does not require the eradication of magnetic
fields to engender an observable effect. 
The experimental concept in this case would be completely different from those 
considered thus far, engendering  $e + n \to {\bar n} +e$, e.g. 
Nuclear stability should also set limits on this source of B-L 
violation~\cite{Berezhiani:2015uya}. 

\section{B-L violation and theories of self-conjugate fermions}
\label{Origin}

In our study of B-L violating operators, 
we have found that it is possible to write down operators which are odd under CPT
but yet are also Lorentz invariant. These 
operators do 
vanish once the anticommuting nature of fermion fields is 
taken into account, though the precise
stature of the results depends on whether the fermion fields are Majorana
or Dirac. In the Majorana case, the demonstration is immediate, following from 
the definition of the Majorana field, Eq.~(\ref{majorcon}), 
and the anticommuting nature of fermion fields, 
whereas in the Dirac case 
it is not. In the latter case canonical quantization and 
a Fourier expansion of the fermion field is required, 
though fermion antisymmetry still plays a crucial role. 
In this section we consider the roots of these differences
and indeed why it should 
be possible to write down a CPT-odd, Lorentz-invariant operator, 
even if it does ultimately vanish. To do this, we recall 
theories of self-conjugate particles with half-integer isospin, 
which are non-local~\cite{Carruthers:prl,HuanLee:prl,fleming:prl,YSJin:plb} 
and have anomalous 
CPT properties~\cite{Kantor,Steinmann,Zumino,Carruthers:jmp1,Carruthers:1968zn,Carruthers:prd3,Carruthers:plb}. 

In attempting to rationalize the spectral pattern of the low-lying, light hadrons, 
Carruthers discovered a class of theories for which the CPT theorem does 
not hold~\cite{Carruthers:prl}. We note 
the pions form a self-conjugate isospin multiplet $(\pi^+, \pi^0, \pi^-)$, 
whereas the kaons form pair-conjugate multiplets 
$(K^+, K^0)$ and $({\bar K}^0, K^-)$, so that the particle and antiparticle appear in 
distinct isospin multiplets. Carruthers discovered that free theories of self-conjugate
bosons with half-integer isospin are nonlocal, that the commutator of two 
self-conjugate fields with opposite isospin components 
do not vanish at 
space-like separations~\cite{Carruthers:prl}, rendering the theory noncausal
 and hence
physically unacceptable. Moreover, since weak local communitivity 
fails~\cite{Carruthers:1968zn}, 
CPT symmetry is no longer expected to hold~\cite{Streater:1989vi}, nor
should the theorem of Ref.~\cite{Greenberg:2002uu} apply. 
These results were quickly generalized, and apply to theories of 
arbitrary spin~\cite{HuanLee:prl,fleming:prl,YSJin:plb}. Consequently 
it is possible to have 
self-conjugate theories of isospin $I=0$, 
but it is not possible to have self-conjugate theories of $I=1/2$. 
These developments are pertinent to the findings in this paper, because 
a Majorana fermion is a self-conjugate particle of $I=0$, whereas the
neutron and antineutron are members of pair-conjugate $I=1/2$ multiplets. 
Since $p-{\bar p}$ oscillations are forbidden by electric charge conservation, 
a theory of $n-{\bar n}$ oscillations need not be a 
theory of self-conjugate isofermions. 
We note, however, that the very 
quark-level operators that 
generate $n-{\bar n}$ oscillations~\cite{Rao:1982gt} would also produce
$p-{\bar p}$ oscillations under the isospin transformation $u \leftrightarrow d$. 
Since QCD is symmetric under $u \leftrightarrow d$ exchange in its chiral limit, 
the admissible B-L violating operators in that case must then necessarily break isospin symmetry,
so that self-conjugate isofermions do not appear. 
Since isospin symmetry is broken in the SM by quark mass and electric charge differences, 
the SM itself is compatible with the appearance of B-L violating operators in the quark sector. 

\section{Summary}
\label{Sum}

In this paper we have determined the restrictions 
on the phases associated 
with the discrete symmetry transformations C, P, and T of fermion fields 
that appear in theories of B-L violation, generalizing the earlier
work of Refs.~\cite{Kayser:1983wm,Kayser:1984ge}. These phase constraints do not impact 
B-L conserving theories because the phases are unimodular, but they 
are key to determining the behavior of B-L violating operators under 
discrete symmetry transformations because they enter as the phase squared. 
As a result, they have important implications for the interplay of 
B-L violating dynamics with the SM. 

We have found that the phase associated with the transformation of a fermion
field under CPT, $\eta_{cpt}$, must always be
imaginary and that the phase associated with P, 
$\eta_p$, must be imaginary for fermions for 
which a P transformation exists.  Generally, however, 
the phase associated with CP, $\eta_{cp}$, 
is indeterminate for B-L violating operators. 
We find that the constraint on $\eta_{cpt}$ 
reconciles the disagreement between
Refs.~\cite{Gardner:2014cma,Berezhiani:2015uya}, to the end that
magnetic fields do indeed quench $n$-${\bar n}$ oscillations mediated
by the operator $\psi^T C \psi + {\rm h.c.}$~\cite{Berezhiani:2015uya}. 
However, spin dependence can still play a key role in $n$-${\bar n}$
transitions, as proposed in Ref.~\cite{Gardner:2014cma}, 
and in this paper we have noted 
the prospects associated with the operator 
$\psi^T C \gamma^\mu \gamma_5 \psi j_\mu + {\rm h.c.}$~\cite{Berezhiani:2015uya}, 
for which $n(+) \rightarrow {\bar n}(-)$, e.g., is mediated by the external current 
$j_\mu$. We note that $n(+)$ and $n(-)$ 
are of the same energy irrespective of the external magnetic fields. 
Moreover, we have shown that 
the appearance of $n$-${\bar n}$ oscillations does not in itself
break CP, in contradistinction to Ref.~\cite{Berezhiani:2015uya}, and  
that this is true irrespective of $\eta_{cp}$.

We expect that CPT is an exact symmetry of 
a local, Lorentz invariant 
quantum field theory~\cite{Streater:1989vi}, 
and if CPT is broken, then
Lorentz invariance fails also~\cite{Greenberg:2002uu}. 
We have found that it is possible to construct B-L violating, Lorentz-invariant 
operators that are either CPT even or odd, but that one set vanishes
due to the anticommuting nature of fermion fields. 
The CPT phase constraint we have found is essential to making 
the nonvanishing B-L operators CPT even. Our ability to prove that
the CPT-odd operators vanish depends on whether the fermion fields
are Majorana or Dirac, with additional assumptions needed in the Dirac case. 
We have explained this in connection to theories of
self-conjugate isofermions, 
for which locality fails~\cite{Carruthers:prl,HuanLee:prl,fleming:prl,YSJin:plb}, 
and the CPT 
properties are 
anomalous~\cite{Kantor,Steinmann,Zumino,Carruthers:jmp1,Carruthers:1968zn,Carruthers:prd3,Carruthers:plb}.  
In this regard Majorana neutrinos and neutrons are distinct, 
because only the latter carry
$I=1/2$. The conservation of electric charge saves 
a theory with $n-{\bar n}$ oscillations, in which $p$-${\bar p}$ oscillations
do not occur, from being a theory of 
self-conjugate isofermions; nevertheless, CPT-odd, Lorentz-invariant operators
can appear, though they ultimately appear to vanish.

\appendix
\section*{Appendices}
\renewcommand{\thesubsection}{\Alph{subsection}}

\subsection{Discrete symmetries --- definitions and other essentials}
\label{CPTdef}

In this appendix we collect the definitions and basic results that underlie 
the central arguments of the paper. 
The discrete-symmetry transformations of a four-component fermion field $\psi(x)$ are given by 
\begin{eqnarray}
&&\mathbf{C}\psi(x)\mathbf{C}^{-1} = \eta_c C \gamma^0 \psi^{\ast}(x) \equiv 
\eta_{c}i\gamma^{2}\psi^{\ast}(x) \equiv \eta_c \psi^c (x) \,,\label{Cdef}\\
&&\mathbf{P}\psi(t,\mathbf{x})\mathbf{P}^{-1}=\eta_{p}\gamma^{0}\psi(t,-\mathbf{x}) \,, 
\label{Pdef}\\
&&\mathbf{T}\psi(t,\mathbf{x})\mathbf{T}^{-1}=\eta_{t}\gamma^{1}\gamma^{3}\psi(-t,\mathbf{x})\,,
\label{Tdef}
\end{eqnarray}
where $\eta_{c}$, $\eta_{p}$, and $\eta_{t}$ are unimodular phase factors of the 
charge-conjugation C, parity P, and time-reversal 
T transformations, respectively, and we have chosen the 
Dirac-Pauli representation for the gamma matrices. Note that 
$\psi^c(x)$ is the conjugate field and that 
$\mathbf{C}^2 \psi(x) \mathbf{C}^{-2}  =\psi(x) $ and $\mathbf{T}^2 \psi(x) \mathbf{T}^{-2}  
= - \psi(x)$, irrespective of arbitrary phases, but that 
$\mathbf{P}^2 \psi(x) \mathbf{P}^{-2}  =\eta_p^2 \psi(x)$. 
Our choices and results conform with those of Ref.~\cite{Gardner:2014cma} if
the arbitrary phases are set to unity --- and 
with those of Ref.~\cite{Branco:1999fs}, though we have chosen
a specific representation of the gamma matrices. 

The plane-wave expansion of a Dirac field $\psi(x)$ is given by
\begin{equation}
\psi(x) = \int \frac{d^3 \mathbf{p}}{(2\pi)^{3/2} \sqrt{2 E}}
\sum_{s=\pm} \left\{ b(\mathbf{p}, s) u(\mathbf{p}, s) e^{-ip\cdot x}
+ d^\dagger(\mathbf{p}, s) v(\mathbf{p}, s) e^{ip\cdot x}
\right\} \,,
\label{free} 
\end{equation}
with spinors defined as 
\begin{equation}
u(\mathbf{p}, s) = {\cal N}
\left(\begin{array}{c}
\chi^{(s)} \\
\frac{\mathbf{\sigma}\cdot \mathbf{p}}{E + M} \chi^{(s)} \\
\end{array}\right) 
\quad ; \quad
v(\mathbf{p}, s) = {\cal N}
\left(\begin{array}{c}
\frac{\mathbf{\sigma}\cdot \mathbf{p}}{E + M} \chi^{\prime\, (s)} \\
\chi^{\prime (s)} \\
\end{array}\right) \,,
\end{equation} 
noting $\chi^{\prime\, (s)} = -i \sigma^2 \chi^{(s)}$, 
$\chi^{+} = \left( \stackrel{1}{{}_0} \right)$,  
$\chi^{-} = \left(\stackrel{0}{{}_1} \right)$, and ${\cal N}=\sqrt{E + M}$. 
Noting that $b (d)$ annihilates a particle (antiparticle), we find the following 
transformation properties: 
\begin{eqnarray}
\textbf{C}b(\textbf{p},s)\textbf{C}^{\dagger}&=& \eta_{c}d(\textbf{p},s)\ ; \ \textbf{C}d^{\dagger}(\textbf{p},s)\textbf{C}^{\dagger}= \eta_{c}b^{\dagger}(\textbf{p},s) \,,\nonumber \\
\textbf{C}b^{\dagger}(\textbf{p},s)\textbf{C}^{\dagger}&=& \eta^{\ast}_{c}d^{\dagger}(\textbf{p},s)\ ; \ \textbf{C}d(\textbf{p},s)\textbf{C}^{\dagger}=\eta^{\ast}_{c}b(\textbf{p},s)\,,  \\
\textbf{P}b(\textbf{p},s)\textbf{P}^{\dagger}&=& \eta_{p}b(-\textbf{p},s)\ ; \ \textbf{P}d^{\dagger}(\textbf{p},s)\textbf{P}^{\dagger}= -\eta_{p}d^{\dagger}(-\textbf{p},s) \,,\nonumber \\
\textbf{P}b^{\dagger}(\textbf{p},s)\textbf{P}^{\dagger}&=& \eta^{\ast}_{p}b^{\dagger}(-\textbf{p},s)\ ; \ \textbf{P}d(\textbf{p},s)\textbf{P}^{\dagger}= -\eta^{\ast}_{p}d(-\textbf{p},s)\,,  \\
\textbf{T}b(\textbf{p},s)\textbf{T}^{-1}&=& s\eta_{t}b(-\textbf{p},-s)\ ;\ \textbf{T}d^{\dagger}(\textbf{p},s)\textbf{T}^{-1}= s\eta_{t}d^{\dagger}(-\textbf{p},-s) \,, \nonumber \\
\textbf{T}b^{\dagger}(\textbf{p},s)\textbf{T}^{-1}&=& s\eta^{\ast}_{t}b^{\dagger}(-\textbf{p},-s)\ ;\ \textbf{T}d(\textbf{p},s)\textbf{T}^{-1}= s\eta^{\ast}_{t}d(-\textbf{p},-s) \,, 
\end{eqnarray} 
where, for convenience, we note that 
\begin{eqnarray}
\gamma^{0} u(\textbf{p},s)= 
u(-\textbf{p},s)\quad &;& \quad
\gamma^{0}v(\textbf{p},s)=-v(-\textbf{p},s) \,, 
\label{Puv}\\
u(\mathbf{p},s)&=&i\gamma^{2}v^{\ast}(\mathbf{p},s) \,,
\label{Cuv} \\
u^\ast(\textbf{p},s) = s \gamma^{1}\gamma^3 u(-\textbf{p},-s)\quad &;& \quad
v^\ast(\textbf{p},s) = s \gamma^{1}\gamma^3 v(-\textbf{p},-s) \,,
\label{Tuv}\\
\gamma^{5}u(\textbf{p},s) &=& -sv(\textbf{p},-s) 
\,.
\label{CPTuv}
\end{eqnarray}

\subsection{Phase restrictions for two-component fields}
\label{CPT2spinor}

In this section we develop the phase restrictions 
associated with the discrete-symmetry transformations 
of two-component Majorana fields. We develop these 
in two different ways: the first by 
connecting 
Dirac fields, and our
earlier phase constraints, with two-component
Majorana fields 
and the second by
analyzing the transformation properties of two-component
Majorana fields directly. 

In Weyl representation, a Dirac spinor can be written as 
\begin{eqnarray}
\psi=\begin{pmatrix}
\xi^{\alpha} \\ \eta_{\dot{\beta}} \end{pmatrix}\,.
\label{Dirac2}
\end{eqnarray}
where $\alpha$ and $\beta$ can be 1 or 2. Here we employ the undotted and dotted 
notation used by Refs.~\cite{Berestetsky:1982aq,Duret:2008zv}. 
The undotted contravariant spinor $\xi^{\alpha}$ and the covariant spinor $\xi_{\alpha}$ 
are in the 
$(\frac{1}{2},0)$ 
representation of the Lorentz group SO(3,1), 
whereas the 
dotted covariant spinor $\eta_{\dot{\beta}}$ and the 
contravariant spinor $\eta^{\dot{\beta}}$ are in 
the 
$(0,\frac{1}{2})$ representation.  
One can raise or lower the undotted indices using the metric of SL(2,C)
\begin{eqnarray}
g_{\alpha\beta}&=&\begin{pmatrix}
0 & 1 \\ -1 & 0 \end{pmatrix}=i\sigma^{2}_{\alpha\beta}, \\
g^{\alpha\beta}&=&\begin{pmatrix}
0 & -1 \\ 1 & 0 \end{pmatrix}=-i\sigma^{2}_{\alpha\beta}\, ,
\end{eqnarray}
i.e., 
\begin{eqnarray}
\xi^{\alpha}=g^{\alpha\beta}\xi_{\beta}=-i\sigma^{2}_{\alpha\beta}\xi_{\beta}\,,
\label{convert}
\end{eqnarray}
and use the same  metric for dotted indices. 

Since the C and P transformations of Eqs.~(\ref{Cdef},\ref{Pdef}) 
connect the $(\frac{1}{2},0)$ and $(0,\frac{1}{2})$ representations of the Lorentz group and
thus the two 
two-component fields in Eq.~(\ref{Dirac2}), a particular two-component field cannot transform
into itself under P or C. However, it can transform into itself under 
CP or CPT (or T)~\cite{Berestetsky:1982aq,Duret:2008zv}, 
so that phase constraints may exist for these particular 
transformations. We will now determine them in two different ways. 

In Sec.~\ref{Major}, we found the phase constraints associated with the 
discrete-symmetry transformations of a Dirac field. 
Revisiting the CP and CPT transformations in Weyl representation, we find 
\begin{eqnarray}
\mathbf{CP}\begin{pmatrix}
\xi^{\alpha}(\mathbf{x},t) \\ \eta_{\dot{\beta}}(\mathbf{x},t) \end{pmatrix} (\mathbf{CP})^{-1}&=&\eta_{cp}i\gamma^{0}\gamma^{2}\begin{pmatrix}
\xi^{\alpha\dagger}(-\mathbf{x},t) \\ \eta_{\dot{\beta}}^{\dagger}(-\mathbf{x},t)\end{pmatrix} 
\,,\\
\mathbf{CPT}\begin{pmatrix}\xi^{\alpha}(x) \\ \eta_{\dot{\beta}}(x) \end{pmatrix} (\mathbf{CPT})^{-1}&=&-\eta_{cpt}\gamma^{5}\begin{pmatrix}
\xi^{\alpha\dagger}(-x) \\ \eta_{\dot{\beta}}^{\dagger}(-x)\end{pmatrix}\,. 
\end{eqnarray}
Since in Weyl representation
\begin{eqnarray}
i\gamma^{0}\gamma^{2}=\begin{pmatrix}
-i\sigma^{2} & 0 \\ 0 & i\sigma^{2}\end{pmatrix};\ \ \ \gamma^{5}=\begin{pmatrix}
-1 & 0 \\ 0 & 1\end{pmatrix}\,
\end{eqnarray}
we use Eq.~(\ref{convert}), e.g., to find 
\begin{eqnarray}
\mathbf{CP} \xi^{\alpha}(\mathbf{x},t)(\mathbf{CP})^{-1}&=&-\eta_{cp}\xi_{\alpha}^{\dagger}(-\mathbf{x},t)\,, \\
\mathbf{CP} \eta_{\dot{\alpha}}(\mathbf{x},t)(\mathbf{CP})^{-1}&=&-\eta_{cp}\eta^{\dot{\alpha}\dagger}(-\mathbf{x},t)\,, \\
\mathbf{CPT} \xi^{\alpha}(x)(\mathbf{CPT})^{-1}&=&\eta_{cpt}\xi^{\alpha\dagger}(-x)\,, \\
\mathbf{CPT} \eta_{\dot{\alpha}}(x)(\mathbf{CPT})^{-1}&=&-\eta_{cpt}\eta_{\dot{\alpha}}^{\dagger}(-x)\,, 
\end{eqnarray} 
where we note, as per Sec.~\ref{Major}, that $\eta_{cpt}\equiv \eta_c\eta_p\eta_t =\pm i$. 
Here we find no direct constraint on the phase $\eta_{cp} \equiv \eta_c \eta_p$, or 
$\eta_t$ for that matter, 
because the analysis of Sec.~\ref{Major} determined that the 
combinations $\eta_{cp}^\ast \lambda$ and $\eta_{t} \lambda$ were imaginary and
real, respectively. Since the phase $\lambda$ has no meaning in the current context, 
no conclusions on $\eta_{cp}$ or $\eta_{t}$ can follow. 

An alternate path to these results comes from the analysis of 
the plane-wave expansion of the two-component
Majorana field $\xi_{a}(x)$~\cite{Case:1957zza,Dreiner:2008tw}:
\begin{eqnarray}
\xi_{\alpha}(x)=\sum_{s}\int \frac{d^{3}\mathbf{p}}{(2\pi)^{3/2}(2E_{\mathbf{p}})^{1/2}}[x_{\alpha}(\mathbf{p},s)a(\mathbf{p},s)e^{-ipx}+\lambda y_{\alpha}(\mathbf{p},s)a^{\dagger}(\mathbf{p},s)e^{ipx}]\,,
\end{eqnarray}
where $x_{\alpha}$ and $y_{\alpha}$ are two-component spinors, whose 
definition and other pertinent details 
can be found in Ref.~\cite{Dreiner:2008tw}. Note that we have included 
a phase factor $\lambda$ in $\xi_{a}(x)$, in analogy to the analysis of Sec.~\ref{Major}. 
It is trivial to check that the phase $\lambda$ included 
here functions in the same way as in Eq.~(\ref{plmajor}) and that it is forced to $1$ when 
$\xi_{\alpha}(x)$ is used to 
constuct a Dirac field~\footnote{Although the notation 
for a Dirac field employed by 
Refs.~\cite{Berestetsky:1982aq,Duret:2008zv} and \cite{Dreiner:2008tw} differs, 
our results are unchanged.}. 
Using the CP transformation of $\xi_{\alpha}(\mathbf{x},t)$~\cite{Berestetsky:1982aq,Duret:2008zv}
\begin{eqnarray}
\mathbf{CP}\xi_{\alpha}(\mathbf{x},t)(\mathbf{CP})^{-1}=\eta_{cp}(\xi^{\alpha})^{\dagger}(-\mathbf{x},t)
\end{eqnarray}
and the  relations~\cite{Dreiner:2008tw} 
\begin{eqnarray}
(x^{\alpha})^{\dagger}(\mathbf{p},s)&=&x^{\dagger\dot{\alpha}}(\mathbf{p},s)=-y_{\alpha}(-\mathbf{p},s)\,, \\
(y^{\alpha})^{\dagger}(\mathbf{p},s)&=&y^{\dagger\dot{\alpha}}(\mathbf{p},s)=x_{\alpha}(-\mathbf{p},s)
\end{eqnarray} 
yield 
\begin{eqnarray}
\mathbf{CP}a(\mathbf{p},s)(\mathbf{CP})^{-1}&=&\eta_{cp}\lambda^{\ast} a(-\mathbf{p},s)\,, \\
\mathbf{CP}a^{\dagger}(\mathbf{p},s)(\mathbf{CP})^{-1}&=&-\eta_{cp}\lambda^{\ast}a^{\dagger}(-\mathbf{p},s)\,. 
\end{eqnarray}
Since $\textbf{CP}$ is a unitary operator, taking the 
Hermitian conjugate of either relation proves that $\eta_{cp}\lambda^{\ast}$ must be 
imaginary. 

Under CPT, we have 
\begin{eqnarray}
\mathbf{CPT}\xi^{\alpha}(x)(\mathbf{CPT})^{-1}=\eta_{cpt}(\xi^{\alpha})^{\dagger}(-x)\,.
\end{eqnarray}
Using the  relations~\cite{Dreiner:2008tw} 
\begin{eqnarray}
x^{\dagger\dot{\alpha}}(\mathbf{p},-s)&=&2sy^{\dagger\dot{\alpha}}(\mathbf{p},s)\,, \\
y^{\dagger\dot{\alpha}}(\mathbf{p},-s)&=&-\frac{1}{2s} x^{\dagger\dot{\alpha}}(\mathbf{p},s)\,, 
\end{eqnarray} 
we find 
\begin{eqnarray}
\mathbf{CPT}a(\mathbf{p},s)(\mathbf{CPT})^{-1}&=& -\frac{1}{2s}\lambda^{\ast}\eta_{cpt}a(\mathbf{p},-s)\,, 
\label{CPT1} \\
\mathbf{CPT}a^{\dagger}(\mathbf{p},s)(\mathbf{CPT})^{-1}&=&2s\lambda\eta_{cpt}a^{\dagger}(\mathbf{p},-s)\,.
\label{CPT2}
\end{eqnarray}
Noting that $\mathbf{CPT}$ is an antiunitary operator, as in Sec.~\ref{Major}, 
we can take the Hermitian conjugate of either equation to show that $\eta_{cpt}$ must be
imaginary. Alternatively, after Ref.~\cite{Kayser:1984ge}, 
we define $\mathbf{CPT}|0\rangle =|0\rangle$ and note 
\begin{eqnarray}
1&=&\langle 0| a(\mathbf{p},s)a^{\dagger}(\mathbf{p},s)|0\rangle \nonumber \\
&=&\langle 0|\mathbf{CPT}a(\mathbf{p},s)\mathbf{CPT}^{-1}\mathbf{CPT}a^{\dagger}(\mathbf{p},s)\mathbf{CPT}^{-1}|0\rangle\,.
\end{eqnarray} 
Then using Eqs.~(\ref{CPT1},\ref{CPT2}) shows that $\eta_{cpt}=\pm i$. 

In summary, we have used two methods to find the phase constraints 
on $\textbf{CP}$ and $\textbf{CPT}$ for two-component fields, and have 
obtained the same results, which are that $\eta_{cp}$ itself is unconstrained, though 
$\eta_{cp}\lambda^\ast$ must be imaginary, and 
$\eta_{cpt}$ is always $\pm i$.

\begin{acknowledgments}
We acknowledge partial support from the Department
of Energy Office of Nuclear Physics under 
contract DE-FG02-96ER40989. The work of SG was performed, in part, 
at the Aspen Center for Physics, which is supported by the National 
Science Foundation grant PHY-1066293, 
and at the Institute of Nuclear Theory (INT). 
We thank M. Voloshin for comments concerning the interaction of 
$n-{\bar n}$ oscillations with electromagnetic fields and A. Vainshtein for 
extensive discussions of this topic and related issues in the context of an 
INT program ``Intersections of BSM Phenomenology and QCD for New Physics Searches''
in the fall of 2015. 
We are also grateful to R. Jaffe for helpful correspondence
regarding the construction of antiquark states in the bag model, and we thank 
A. Nelson for a discussion concerning parity and two-component Majorana spinors, occurring within 
the context of the aforementioned INT program. We also thank T.~Goldman 
for bringing early work on phase restrictions in discrete symmetry transformations 
to our attention.

\end{acknowledgments}
\bibliographystyle{doiplain}

\end{document}